\documentclass[fleqn,10pt]{wlscirep}
\title{Security enhanced memory for quantum state}

\author[1,*]{Tetsuya Mukai}
\affil[1]{NTT Basic Research Laboratories, NTT Corporation, 3-1, Morinosato-Wakamiya, Atsugi, Kanagawa 243-0198, Japan}

\affil[*]{mukai.tetsuya@lab.ntt.co.jp}

\begin{abstract}
Security enhancement is important in terms of both classical and quantum information. 
The recent development of a quantum storage device is noteworthy, and a coherence time of one second or longer has been demonstrated. On the other hand, although the encryption of a quantum bit or quantum memory has been proposed theoretically, no experiment has yet been carried out. 
Here we report the demonstration of a quantum memory with an encryption function that is realized by scrambling and retrieving the recorded quantum phase. We developed two independent Ramsey interferometers on an atomic ensemble trapped below a persistent supercurrent atom chip. By operating the two interferometers with random phases, the quantum phase recorded by a pulse of the first interferometer was modulated by the second interferometer pulse. The scrambled quantum phase was restored by employing another pulse of the second interferometer with a specific time delay. This technique paves way for improving the security of quantum information technology.
\end{abstract}
\begin{document}

\flushbottom
\maketitle
\thispagestyle{empty}

\section*{Introduction}

A quantum memory, namely a device that records a non-classical state, is an indispensable resource for advanced quantum information processing. Many quantum memory studies have already been reported that use various schemes and hardware \cite{Harber_2002, Treutlein_2004, Dutt_2007, Morton_2008, Lvovsky_2009, Clausen_2011, Saglamyurek_2011, Nicolas_2014}. In these studies, an atomic ensemble has been extensively investigated because of its good accessibility with photonic flying qubits and its relatively long coherence time thanks to its isolation from the environment. For example, the single spin excitation of an atomic ensemble is used as a photonic state memory for a quantum network  \cite{Sangouard_2011}, and a Rabi flop and a Ramsey interferometer can be regarded as a quantum bit or quantum memory for quantum superposition \cite{Harber_2002, Treutlein_2004}. With a Ramsey interferometer, a coherence time of several seconds has been demonstrated with an atom chip technique \cite{Deutsch_2010, Bernon_2013}, and the application of a long coherence time has been possible with state-of-the-art technology. 

Quantum information is intrinsically secure because of the measurement back-action and the no-cloning theorem. However, the quantum information preserved in a quantum memory with a single key lock is in peril of leaking as a result of key theft and direct access to the memory. 
To increase the security, the encryption of a quantum bit or quantum memory has been proposed theoretically \cite{Boykin_2003}. However, no experiment has yet been carried out.
Here we report the experimental demonstration of a double-key-lock quantum memory by encrypting the quantum phase recorded on an atomic ensemble trapped below a persistent supercurrent atom chip.

\section*{Results}
\subsection*{Ramsey interferometry as a single-key-lock quantum memory}
\begin{figure*}[thbp]
\begin{center}
\includegraphics[width=16cm]{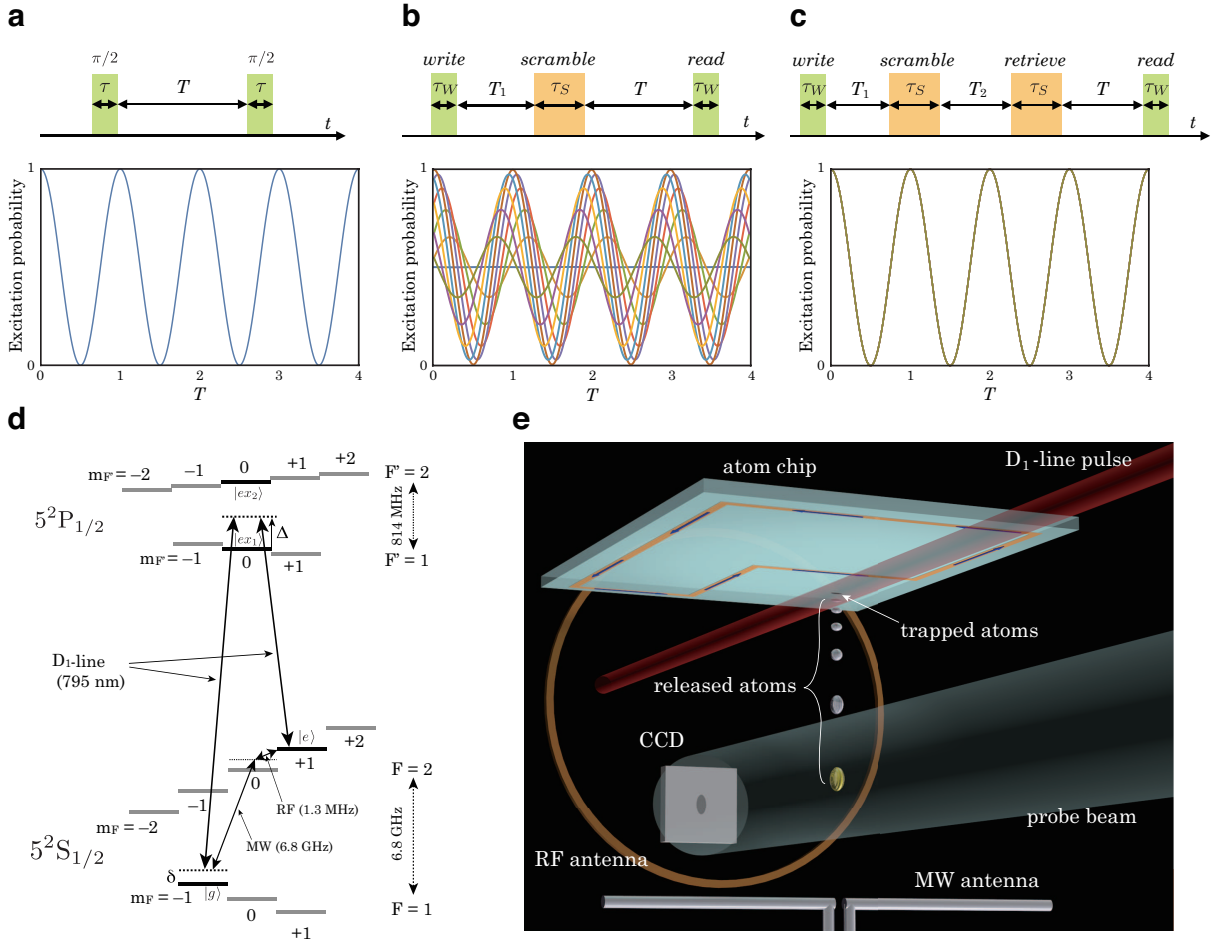}
\caption{{\bf Security enhanced quantum memory}. The horizontal axis $T$ of the plot in {\bf a, b}, and {\bf c} is normalized with $2\pi/\delta_{j}$. {\bf a}, Pulse sequence of the Ramsey interferometry and the calculated excitation probability of the Ramsey flop. {\bf b}, Scrambling pulse sequence and the calculated excitation probability. The excitation probability includes several curves calculated with a phase difference $\phi_{S} \in [0,2\pi]$. {\bf c}, Scramble-retrieve pulse sequence and the calculated excitation probability. The excitation probability includes several curves with a phase difference $\phi_{S} \in [0,2\pi]$. {\bf d}, Relevant energy level for the stimulated Raman transition using D$_{1}$-line / MW-RF fields. {\bf e}, Illustration of the experimental setup for security enhanced quantum memory. D$_{1}$-line, MW, and RF fields are applied when atoms are trapped in a magnetic potential below the atom chip. Excitation probability is measured after a 16~ms TOF with state selective absorption measurements.}
\label{Ramsey_flop}
\end{center}
\end{figure*}

The encryption process requires unitary operations for reliable decryption, and the Ramsey interferometry technique is available for this purpose. 
Ramsey interferometry was originally designed for measuring the resonant frequency of a molecular beam with two interacting fields that are applied at geometrically separated points to resolve small frequency splitting \cite{Ramsey_1949, Ramsey_1950}. After the invention of a laser cooling and trapping technique for atoms, the molecular beam was replaced by a trapped atomic cloud, and the geometrically separated fields were replaced by temporally separated field pulses. Here we assume a two-level atom with a ground state $|g\rangle$ and an excited state $|e\rangle$ separated by an energy difference $E = \hbar \omega_{a}$. The interaction field with frequency $\omega_{j}$ resonantly couples the transition with a small frequency detuning $\delta_{j} = \omega_{j} - \omega_{a}$. The field amplitude is represented with the Rabi frequency $\Omega_{j}$, and the practical Rabi frequency is described as $\tilde{\Omega}_{j} = \sqrt{\Omega_{j}^{2} + \delta_{j}^{2}}$. The interaction of the field with atoms causes Rabi oscillation between the ground $|g\rangle$ and the excited $|e\rangle$ states. The field interaction time $\tau_{j}$ is set to meet the $\pi/2$ pulse condition, i.e. $\Omega_{j} \tau_{j} = \pi/2$, so as to transfer half of the population in each ground or excited state. The two $\pi/2$ pulse interactions are temporally separated with a time interval $T$. After using a rotating wave approximation ($\omega_{j} \sim \omega_{a}$) and the assumption of a short pulse duration $\tau_{j}$ relative to the time interval $T$ ($\tau_{j}/T \ll1$), the time evolution of the wave function $| \Psi(t) \rangle$ is described with the following equation,
\begin{equation}
| \Psi(T) \rangle = U_{j}(\omega_{j} T + \phi_{0})~E(T)~U_{j}(\phi_{0})~| \Psi(0) \rangle,
\label{Ramsey flops matrix}
\end{equation}
where $\phi_{0}$ represents an initial phase, which we can set at zero without loss of generality, and $U_{j}(\phi)$ and $E(t)$ represent the following field interaction and time evolution in the base $\{ |g\rangle, |e\rangle\}$;
\begin{eqnarray*}
U_{j}(\phi) &=& 
\begin{bmatrix}
	e^{i \frac{\delta_{j}}{2} \tau_{j}}\{\cos(\frac{\tilde{\Omega}_{j}}{2} \tau_{j} ) - i\frac{\delta_{j}}{\tilde{\Omega}_{j}}\sin(\frac{\tilde{\Omega}_{j}}{2} \tau_{j})\} & 
	-ie^{i (\frac{\delta_{j}}{2} \tau_{j} + \phi)}\frac{\Omega_{j}}{\tilde{\Omega}_{j}}\sin(\frac{\tilde{\Omega}_{j}}{2} \tau_{j}) \\
	-ie^{-i (\frac{\delta_{j}}{2} \tau_{j} + \phi)}\frac{\Omega_{j}}{\tilde{\Omega}_{j}}\sin(\frac{\tilde{\Omega}_{j}}{2} \tau_{j}) & 
	e^{-i \frac{\delta_{j}}{2} \tau_{j}}\{\cos(\frac{\tilde{\Omega}_{j}}{2} \tau_{j} ) + i\frac{\delta_{j}}{\tilde{\Omega}_{j}}\sin(\frac{\tilde{\Omega}_{j}}{2} \tau_{j})\}
\end{bmatrix}\\
E(t) &=&
\begin{bmatrix}
	e^{i \omega_{a} t/2} & 0 \\
	0 & 
	e^{-i \omega_{a} t/2}
\end{bmatrix}.
\end{eqnarray*}
Starting with the atomic ground state, i.e. $|g\rangle = 1$ and $|e\rangle = 0$, the excitation probability $P_{e}(t)$ can be approximately written as follows,
\begin{equation}
P_{e}(T) = 4 \frac{\Omega_{j}^{2}}{\tilde{\Omega}_{j}^{2}}\sin^{2}{\left( \frac{\tilde{\Omega}_{j}\tau_{j}}{2}\right)}\left[\cos \left( \frac{\delta_{j} T}{2} \right) \cos\left(\frac{\tilde{\Omega}_{j}\tau_{j}}{2}\right) -\frac{\delta_{j}}{\tilde{\Omega}_{j}} \sin \left( \frac{\delta_{j} T}{2} \right) \sin \left(\frac{\tilde{\Omega}_{j}\tau_{j}}{2}\right)\right]^{2}.
\label{Excitation probability}
\end{equation}
When the Rabi frequency $\Omega_{j}$ and the pulse duration time $\tau_{j}$ are fixed, the excitation probability exhibits a cosine curve as a function of the time interval $T$ with a frequency equal to the detuning $\delta_{j}/2\pi$ as shown in Fig.~\ref{Ramsey_flop}a.

A Ramsey interferometer can be regarded as a single-key-lock quantum memory that records quantum superposition. For recording arbitrary superposition of $|g\rangle$ and $|e\rangle$ states, the first pulse area should be changed from $\pi/2$ to a suitable one. In the readout process, the second pulse with the same pulse area as the first one should be shined at the time when the interferometer phase evolves to be $(2k+1)\pi$, where $k$ is an integer. If we can measure 100 \% distribution in $|g\rangle$ state, we can confirm that we have recorded the superposition of $|g\rangle$ and $|e\rangle$ states defined by the first pulse area. The phase, frequency, and application time of the first pulse constitute crucial information for reading out the recorded quantum superposition, and from this point of view, the information of the first pulse is a ``key" as regards the quantum memory for readout. To simplify the following description of encryption technique, we employ the standard Ramsey interferometer setting with $\pi/2$ pulse as the first and second pulses in this article. 

\subsection*{Double-key-locking scheme}
To enhance the security of the above mentioned single-key-lock quantum memory, we propose encrypting the stored quantum phase information. For this purpose, we introduce another independent Ramsey interferometer that is operated with a field that has a random phase difference with respect to the field of the first Ramsey interferometer. To avoid confusion, we call the first and second Ramsey interferometers a write-read interferometer (WRI) and a scramble-retrieve interferometer (SRI), respectively. In addition, we call the first and second pulses of the WRI (SRI) {\it write} and {\it read} ({\it scramble} and {\it retrieve}) pulses, respectively. 

The operation sequence of the phase encryption is constituted with the following pulse applications. First, a {\it write} pulse is used to initialize and record the quantum phase. Then, a {\it scramble} pulse is applied after a time interval $T_{1}$ from the {\it write} pulse. Finally, a {\it read} pulse is applied with a time interval $T$ from the application of the {\it scramble} pulse to prepare for the excitation probability readout. The time evolution of the wave function $| \Psi_{s}(t) \rangle$ of this phase scramble sequence can be calculated with the following equation,
\begin{eqnarray}
| \Psi_{s}(T) \rangle &=& U_{W}(\omega_{W}(T+T_{1}))~E(T)~U_{S}(\omega_{S} T_{1}+\phi_{S})~E(T_{1})~U_{W}(0)~| \Psi(0) \rangle,
\label{Scrambled Ramsey flops}
\end{eqnarray}
where $\phi_{S}$ shows the relative phase difference between the field of the SRI with respect to the field of the WRI. The field interaction $U_{j}(\phi)$ and frequency $\omega_{j}$ with $j = W$ or $S$ represent the interaction matrix and the pulse-field frequency of WRI ($j=W$) and SRI ($j=S$), respectively. The excitation probability after this sequence is plotted in Fig.~\ref{Ramsey_flop}b. After the application of a {\it scramble} pulse, the recorded quantum phase depends on the phase difference $\phi_{S}$. By setting a completely random relative phase difference $\phi_{S}$, we can introduce $\pi$ phase ambiguity into the recorded quantum phase, and the readout after the application of the {\it read} pulse fails to reconstruct a clear Ramsey flop.

To retrieve the quantum phase, a {\it retrieve} pulse is employed after the application of a {\it scramble} pulse with a time interval $T_{2}$ that has a value equal to $(2n+1)\pi/\delta_{S}$, where $n$ is an integer and $\delta_{S}$ represents the Raman detuning of the SRI. With this time interval the state flipped by the {\it scramble} pulse is flipped back by the {\it retrieve} pulse after a $\pi$ phase evolution. The time evolution of the wavefunction $| \Psi_{r}(t) \rangle$ after this sequence can be calculated as follows, 
\begin{eqnarray}
| \Psi_{r}(T) \rangle &=& U_{W}(\omega_{W}(T + T_{1} + T_{2}))~E(T)~U_{S}(\omega_{S} (T_{1} + T_{2})+\phi_{S})~E(T_{2})~U_{S}(\omega_{S} T_{1}+\phi_{S})~E(T_{1})~U_{W}(0)~| \Psi(0) \rangle.
\label{Retrieved Ramsey flops}
\end{eqnarray}
The calculated excitation probability is plotted in Fig.~\ref{Ramsey_flop}c. As expected, the Ramsey flop is recovered by employing the {\it retrieve} pulse. 

The scramble-retrieve sequence can be regarded as a double-key-lock quantum memory operation. The first key is the {\it write} pulse information and the second key is the {\it scramble} pulse information. Without information about the phase, frequency, and application time of the two pulses, the recorded quantum phase is difficult to recover with high accuracy.

\subsection*{Experimental setup and scrambling-retrieving measurements}
The experiment was performed with $^{87}{\rm Rb}$ atoms as follows. We defined the spin states $| F, m_{\rm F} \rangle = | 1, -1 \rangle$ and $| 2, 1 \rangle$ of the ground state ${\rm 5 ^{2}S_{1/2}}$ as $| g \rangle$ and $| e \rangle$, and the $| 1, 0 \rangle$ and $| 2, 0 \rangle$ of the excited state ${\rm 5 ^{2}P_{1/2}}$ as $| ex_{1} \rangle$ and $| ex_{2} \rangle$, respectively. The atomic cloud, initially trapped in a magneto-optical trap (MOT), was optically pumped to the spin state $| g \rangle$, and transferred to the magnetic potential of a persistent supercurrent atom chip \cite{Mukai_2007}. After cooling by radio frequency (RF) forced evaporation, a Bose-Einstein condensate (BEC) was obtained $100~{\rm \mu m}$ below the chip surface (see Methods) \cite{ChipBEC_2014}. 

\begin{figure}[thbp]
\begin{center}
\includegraphics[width=8.5cm]{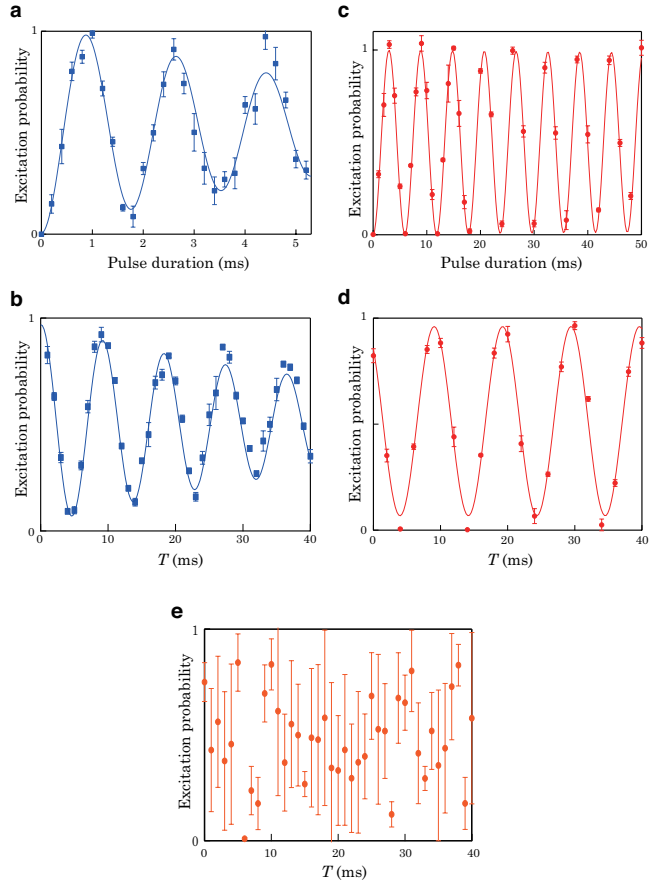}
\caption{{\bf Rabi and Ramsey flops driven by D$_{1}$-line / MW-RF fields.} The lines are damped sinusoidal fit to the data. {\bf a}, D$_{1}$-line Rabi flop. {\bf b}, D$_{1}$-line Ramsey flop. {\bf c}, MW-RF Rabi flop. {\bf d}, MW-RF Ramsey flop. {\bf e}, Relative phase difference between the D$_{1}$-line and MW-RF fields obtained with a Ramsey interferometry using the MW-RF and D$_{1}$-line fields as the first and second $\pi/2$ pulses, respectively. The error bars reflect 1$\sigma$ statistical uncertainty.}
\label{Rabi_Ramsey}
\end{center}
\end{figure}

The WRI was constituted with the D$_{1}$-line stimulated Raman transition that coupled the $| g \rangle$ and $| e \rangle$ spin states via an intermediate state as illustrated in Fig.~\ref{Ramsey_flop}d (see Methods). In our experiment a 565~Hz (= $\Omega_{W}/2\pi$) Rabi flop was observed (Fig.~\ref{Rabi_Ramsey}a), and the $1/e$ amplitude coherence time of the Ramsey flop reached more than 30 times larger than a cycle of the Rabi flop (Fig.~\ref{Rabi_Ramsey}b). 

The SRI was constituted with a microwave (MW) and a radio-frequency (RF) field that coupled the $| g \rangle$ and $| e \rangle$ spin states via two photon transition (see Methods). The MW field was irradiated from a dipole antenna located below the atomic cloud, and the RF field was applied from a loop antenna attached beside the atom chip (Fig.~\ref{Ramsey_flop}e). With this setup a 169~Hz (= $\Omega_{S}/2\pi$) Rabi flop was observed (Fig.~\ref{Rabi_Ramsey}c), and the $1/e$ amplitude coherence time of the Ramsey flop was more than 100 times larger than a Rabi flop cycle (Fig.~\ref{Rabi_Ramsey}d).

The excitation probability of the atomic cloud was estimated by measuring the number of atoms in the $| g \rangle$ and $| e \rangle$ states with absorption images of atomic density taken after a time-of-flight (TOF) of 16~ms (see Methods).

\begin{figure}[thbp]
\begin{center}
\includegraphics[width=7cm]{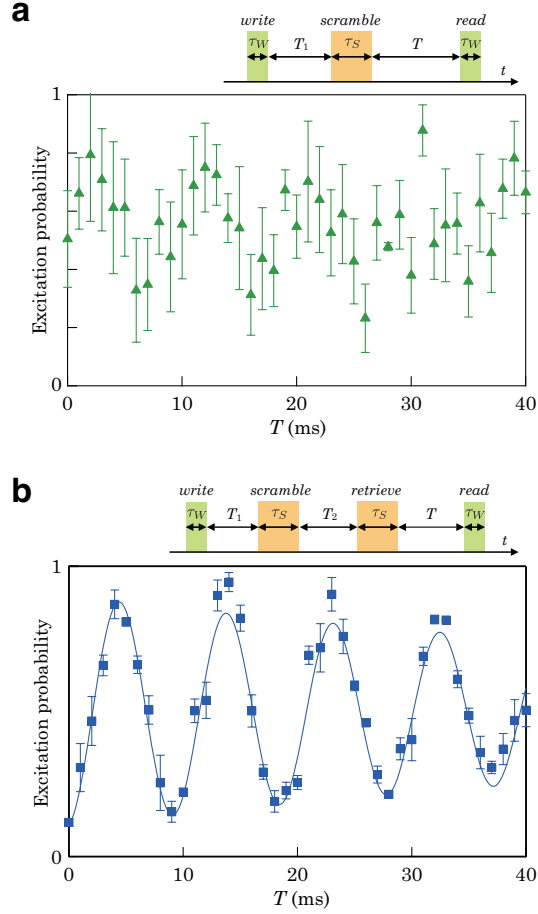}
\caption{{\bf Experimentally obtained scrambled and retrieved Ramsey flops.} {\bf a} Encrypted state: excitation probability after the application of a {\it scramble} pulse in between the {\it write} and {\it read} pulses. The large error bars indicate the fluctuation of the excitation probability caused by the random phase modulation of the {\it scramble} pulse. {\bf b} Decrypted state: excitation probability after the application of {\it scramble} and {\it retrieve} pulses in between the {\it write} and {\it read} pulses. After the application of a {\it retrieve} pulse, the Ramsey flop is restored. The line is a damped sinusoidal fit to the data, and the error bars reflect 1$\sigma$ statistical uncertainty.}
\label{Scramble_Retrieve}
\end{center}
\end{figure}
\begin{table}[ht]
\centering
\begin{tabular}{|l|l|l|}
\hline
parameter & value\\
\hline
WRI field Rabi frequency & $\Omega_{W}$ = $565 \times 2\pi$~Hz\\
WRI pulse duration time & $\tau_{W}$ = 0.44~ms\\
WRI Raman detuning & $\delta_{W}$ = $110\times 2\pi$~Hz\\
SRI field Rabi frequency & $\Omega_{S}$ = $169 \times 2\pi$~Hz\\
SRI pulse duration time & $\tau_{S}$ = 1.48~ms\\
SRI Raman detuning & $\delta_{S}$ = $100\times 2\pi$~Hz\\
{\it write}-{\it scramble} time interval & $T_{1}$ = 5~ms\\
{\it scramble}-{\it retrieve} time interval & $T_{2}$ = 5~ms\\
\hline
\end{tabular}
\caption{\label{tab:table1}Parameters used in the experiment}
\end{table}

Figure~\ref{Scramble_Retrieve} shows an excitation probability obtained with the quantum phase scramble-retrieve experiments. The parameters used in the measurement are summarized in Table~\ref{tab:table1}.
In Fig.~\ref{Scramble_Retrieve}a the excitation probability fluctuated significantly after the application of a {\it scramble} pulse. Although the maximum phase modulation was limited to $\pi$, a large ambiguity was introduced and it is not practically possible to identify the quantum phase. On the other hand, in Fig.~\ref{Scramble_Retrieve}b, a clear Rabi flop was restored by the application of a {\it retrieve} pulse.

\section*{Discussion}

The crux of the scramble-retrieve scheme is the random phase difference between the WRI and SRI fields. In our experiment, the random phase difference was obtained by employing D$_{1}$-line and MW-RF fields, which have five orders of difference in frequency (see Methods). On the other hand, given the fact that the utilization of two fields with many orders of difference in frequency is not essential for generating a random phase difference, the scramble-retrieve scheme must be possible with only D$_{1}$-line or MW-RF fields when the phase difference is randomized. However, the demonstration of coherent control using both D$_{1}$-line and MW-RF fields promises a wide range of applications. One of the practical applications increases the scalability by taking advantage of the difference in spatial resolution. By trapping more than two atomic clouds on a chip, individual and global control is possible with D$_{1}$-line and MW-RF fields, respectively.

\begin{figure}[thbp]
\begin{center}
\includegraphics[width=7.0cm]{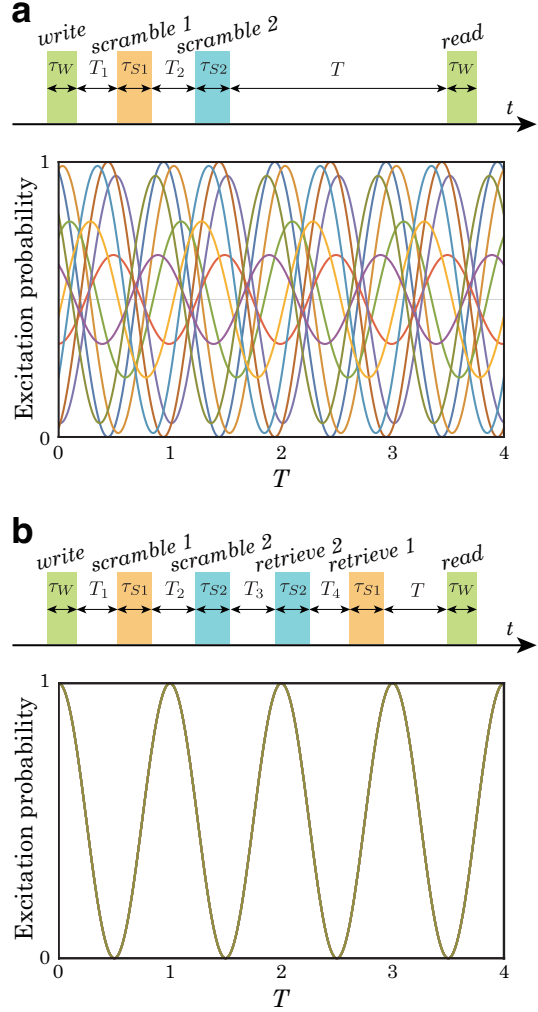}
\caption{{\bf Double scrambling encryption and decryption.} The horizontal axis $T$ of the plot in {\bf a} and {\bf b} is normalized with $2\pi/\delta_{j}$. The excitation probability includes several curves calculated with a phase difference $\phi_{S1} \in [0,2\pi]$ and $\phi_{S2}=\pi/15$. {\bf a}, Double scrambling pulse sequence and calculated excitation probability. The phase ambiguity of excitation probability is expanded to $\sim2\pi$ with the double scrambling and it is completely impossible to determine the quantum phase. {\bf b}, Retrieve sequence for the double scrambling encryption and the excitation probability. The {\it retrieve}-1 pulse for the {\it scramble}-1 pulse comes after the {\it retrieve}-2 pulse for the {\it scramble}-2 pulse. The pulse application time should satisfy $T_{2}+T_{3}+T_{4} +2 \tau_{S2}= (2m+1)\pi/\delta_{S1}$ and $T_{3} = (2n+1)\pi/\delta_{S2}$, where $m$ and $n$ are integers. The $\delta_{j}$, $\phi_{j}$, and $\tau_{j}$ with $j = S1$ or $S2$ represents the detuning, phase difference, and pulse duration of the SRI-1 ($j = S1$) and SRI-2 ($j=S2$), respectively, and the $T_{k}$ with $k = $~1, 2, 3, and 4 represents the time interval between the {\it write} and {\it scramble}-1 pulses ($k=1$), the {\it scramble}-1 and {\it scramble}-2 pulses ($k=2$), the {\it scramble}-2 and {\it retrieve}-2 pulses ($k=3$), and the {\it retrieve}-2 and {\it retrieve}-1 pulses ($k=4$), respectively. }
\label{6pulse}
\end{center}
\end{figure}

To increase the ambiguity of the scrambled quantum phase, a double scramble scheme, i.e. triple Ramsey interferometers, works as a straightforward extension of the double Ramsey interferometers scheme (see Methods). Figure~\ref{6pulse} shows the excitation probability calculation for the double scramble scheme. With this sequence, the quantum phase is sufficiently scrambled and it is completely impossible to determine the quantum phase.

The other interesting application of this scheme is secret sharing with the double-key-lock quantum memory. When Alice records data in a quantum memory with her {\it write} pulse and Bob encrypts it with his {\it scramble} pulse, then Alice and Bob cannot read out the recorded data without their cooperation.

In summary, we have achieved a security enhanced quantum memory with a double Ramsey interferometry technique. The demonstration of coherent phase control using both a D$_{1}$-line laser and MW-RF fields reveals the potential for a wide range of applications connecting a photonic network and microwave devices. In addition, the multiple-key-lock quantum memory is applicable as a secret sharing protocol for enhancing information security.

\section*{Methods}
\subsection*{Preparation of a BEC below a persistent supercurrent atom chip}
The preparation of an atomic cloud or a Bose-Einstein condensate below a persistent supercurrent atom chip is detailed in our previous reports~\cite{Mukai_2007, ChipBEC_2014}. Although we were able to obtain more than $2 \times 10^{5}$ atoms in a pure condensate, to avoid the loss of coherence for the collisional interactions, the number of atoms in a condensate was reduced to $5 \times 10^{4}$ (atomic density = $1\times10^{13}$~cm$^{-3}$) by reducing the MOT loading time.
The magnetic field of the potential bottom was adjusted to $B_{min} = 0.32$~mT to meet the magic $B$ field condition. In a magic $B$ field, the spin states $| g \rangle$ and $| e \rangle$, which were employed as the $| g \rangle$ and $| e \rangle$ states of the Ramsey interferometer, have the same magnetic field dependence up to the second order of the Zeeman shift \cite{Harber_2002}.

\subsection*{D$_{1}$-line stimulated Raman transition}
The WRI was constituted with the D$_{1}$-line stimulated Raman transition that coupled the $| g \rangle$ and $| e \rangle$ spin states via an intermediate state, which was detuned by $\Delta$ from the excited state $| ex_{1} \rangle$ as illustrated in Fig~\ref{Ramsey_flop}d. By setting the detuning $\Delta$ close to half the value of the energy difference of the $| ex_{1} \rangle$ and $| ex_{2} \rangle$ states, the energy shift of the Rabi flop reached close to zero, which made the Raman detuning less sensitive to the interacting field intensity. The pump and the Stokes field were generated from a diode laser and an electro-optic modulator driven with a frequency detuned from the energy difference between $| g \rangle$ and $| e \rangle$ by $\delta_{W}/2\pi$. The frequency of the diode laser was stabilized to give it a linewidth of less than 1~kHz with an ultra low expansion (ULE) cavity. The pump and the Stokes fields were linearly polarized perpendicular to the principal axis of the trapped condensate to excite $\sigma^{+}$ and $\sigma^{-}$ transitions.

\subsection*{MW-RF Raman transition}
The SRI consisted of a microwave (MW) and a radio frequency (RF) field that coupled the $| g \rangle$ and $| e \rangle$ spin states via a two-photon transition. The MW field at frequency $f_{\rm MW} = 6.833378$~GHz and the RF field at frequency $f_{\rm RF} = 1.3$~MHz were generated by commercial synthesizers phase locked to a highly stable 10~MHz quartz oscillator. The sum of the MW and RF frequency was detuned from the energy difference between $| g \rangle$ and $| e \rangle$ by $\delta_{S}/2\pi$.

\subsection*{Measurement of excitation probability}
The excitation probability was estimated by measuring the number of atoms in the $| g \rangle$ and $| e \rangle$ states with an absorption image of atomic density taken after a 16~ms time-of-flight (TOF). To image the combined number of atoms in the $| g \rangle$ and $| e \rangle$ states, a short repumping pulse, which was resonant with the D$_{2}$-line ($F$ = 1 to $F'$ = 2) transition, pumped the $| g \rangle$ state into the $F$ = 2 manifold. About 100~$\mu$s later, a $\sigma^{+}$ polarized probe beam, which was resonant with the D$_{2}$-line ($F$ = 2 to $F'$ = 3) transition, was illuminated to obtain an image of the shadow of the atomic cloud with a charge-coupled device (CCD) camera. The image of the $| e \rangle$ state was obtained with the probe beam without applying the repumping pulse. We performed five measurements with the same setting parameters, and averaged the results to obtain a single data point with a standard deviation.

\subsection*{Random phase difference}
The random phase difference between the D$_{1}$-line and MW-RF fields was caused by the spontaneous emission of the D$_{1}$-line laser. The phase variance $\langle \Delta \phi^{2} \rangle$ is related to the laser linewidth $\Delta \omega$ with an equation $d\langle \Delta \phi^{2} \rangle/ dt = \Delta \omega$ \cite{Henry1982}. In our experiment the phase variance was sufficiently large by employing the laser linewidth of 1~kHz and the measurement time interval of 47 seconds for taking one absorption image of an atomic cloud. Figure~\ref{Rabi_Ramsey}e shows the excitation probability obtained with a Ramsey interferometry using the MW-RF and D$_{1}$-line fields as the first and second $\pi/2$ pulses, respectively. The relative phase difference is sufficiently large, and we recognized that the D$_{1}$-line and MW-RF fields have a random phase difference.

\subsection*{Double scramble scheme: triple Ramsey interferometers}
A double scramble scheme is a straightforward extension of the scramble-retrieve scheme described above. To distinguish two scramble-retrieve interferometers, we name them SRI-1 and SRI-2, and {\it scramble}-i and {\it retrieve}-i with i = 1 or 2, the {\it scramble} and {\it retrieve} pulses of SRI-1 (i = 1) and SRI-2 (i = 2), respectively. After the {\it write} and {\it scramble}-1 pulses application, the {\it scramble}-2 pulse is applied. The {\it retrieve}-1 pulse comes after the {\it retrieve}-2 pulse. The pulse application time should satisfy $T_{2}+T_{3}+T_{4} +2 \tau_{S2}= (2m+1)\pi/\delta_{S1}$ and $T_{3} = (2n+1)\pi/\delta_{S2}$, where $m$ and $n$ are integers. The $\delta_{j}$, $\phi_{j}$, and $\tau_{j}$ with $j = S1$ or $S2$ represents the detuning, phase difference, and pulse duration of SRI-1 ($j = S1$) and SRI-2 ($j=S2$), respectively. The $T_{k}$ with $k =$~1, 2, 3, or 4 represents the time interval between the {\it write} and {\it scramble}-1 pulses ($k=1$), the {\it scramble}-1 and {\it scramble}-2 pulses ($k=2$), the {\it scramble}-2 and {\it retrieve}-2 pulses ($k=3$), and the {\it retrieve}-2 and {\it retrieve}-1 pulses ($k=4$), respectively.

\section*{Acknowledgements}

This work received support from the Japan Society for the Promotion of Science (JSPS) through its KAKENHI Grant Number JP 26286068 and the Japan Science and Technology Agency (JST) through its Core Research for Evolutional Science and Technology (CREST) Grant Number JPMJCR1671. The author thanks N. Imoto for useful comments and conversations.

\section*{Author contributions statement}

T.M. conceived and conducted the experiment, analyzed the results, and wrote the manuscript.

\end{document}